\documentclass[twocolumn,bibnotes,aps,prl,floatfix]{revtex4}
\usepackage{graphicx}
\usepackage{amsmath}
\usepackage{latexsym}
\usepackage{amsfonts}
\usepackage{amssymb}
\usepackage{array}
\usepackage{color}
\usepackage{mathtools}
\usepackage{subfigure}
\usepackage{xcolor}
\usepackage{verbatim}
\usepackage{gensymb}
\usepackage[bb=boondox]{mathalfa}
\usepackage{bigstrut}
\usepackage{array}
\usepackage{tabularx}
\usepackage{booktabs}

\usepackage{comment}
\usepackage{physics}
\usepackage{float}

\begin{document}
\newcommand{\Q}[1]{{\color{red}#1}}
\newcommand{\older}[1]{{\color{gray}#1}}
\newcommand{\red}[1]{{\color{black}#1}}
\newcommand{\green}[1]{{\color{green}#1}}
\newcommand{\update}[1]{{\color{magenta}#1}}
\newcommand{\Change}[1]{{\color{green}#1}}
\newcommand{\T}{\mathrm{Tr}}

\title{Nullifiers of non-Gaussian cluster states through homodyne measurement}
\author{Vojt\v{e}ch Kala}
\email{kala@optics.upol.cz}
\affiliation{Centre for Quantum Information and Communication, École polytechnique de Bruxelles, CP 165, Université libre de Bruxelles, 1050 Brussels, Belgium}
\affiliation{Department of Optics, Palack\'y University, 17. listopadu 1192/12, 77146 Olomouc, Czech Republic}

\author{Casper A. Breum}
\author{Mikkel V. Larsen}
\author{Ulrik L. Andersen}
\author{Jonas S. Neergaard-Nielsen}
\affiliation{Center for Macroscopic Quantum States (bigQ), Department of Physics, Technical University of Denmark,
Building 307, Fysikvej, 2800 Kongens Lyngby, Denmark}
\author{Radim Filip}

\author{Petr Marek}

\affiliation{Department of Optics, Palack\'y University, 17. listopadu 1192/12, 77146 Olomouc, Czech Republic}

\begin{abstract}
In continuous variable optical platforms, large-scale Gaussian cluster states have already been demonstrated, but non-Gaussian resources are essential to achieve universality and fault tolerance in measurement-based quantum computation. However, characterizing and certifying non-Gaussian cluster states remains an outstanding challenge. Here, we introduce a general framework for the characterization of non-Gaussian cluster states based on non-Gaussian nullifiers, extending the widely used Gaussian nullifier concept. We show that these nullifiers can be directly evaluated from homodyne measurement data, making them experimentally accessible. As an illustration, we derive and experimentally demonstrate non-Gaussian nullifiers for photon-subtracted squeezed states. Our results provide a practical and operational tool for certifying quantum non-Gaussianity in large-scale optical cluster states.
\end{abstract}

\maketitle
The main advantage of quantum optics as a platform for continuous variable quantum computing \cite{Menicucci,LarsenFT,AghaeeRad2025} is their intrinsic scalability in the number of bosonic modes. Deterministic entanglement generation via Gaussian operations enables the feasible preparation of enormously large entangled states, known as cluster states \cite{Loock2007}, which have already been demonstrated multiple times \cite{Larsen,Asavanant,Yoshikawa}. The remaining key challenge to unlock universality and fault tolerance on this platform is the preparation of non-Gaussian states with sufficiently high quality and complexity, and their integration into cluster state \cite{Gu,KonnoGKP,Kalatelep}.

The experimental preparation of continuous variable Gaussian cluster states consists of producing a set of squeezed states that are entangled via beam splitter operations \cite{Larsen,Asavanant}. Individual modes of the generated cluster states are eventually measured via homodyne detection, which naturally provides access to their quantum correlations and enables their characterization. Squeezing of noise in mutually commuting operators, called nullifiers, shows that the experimental implementation approaches the ideal cluster state \cite{Loock2007}. An ideal cluster state is an eigenstate of the nullifiers, resulting in vanishing noise. The experimental version, burdened by decoherence and finite squeezing, is then certified when its nullifier noise falls below a given threshold \cite{Larsen}.

\red{On the other hand, experimental preparation of non-Gaussian states is far more challenging as their non-Gaussian features are typically more fragile than Gaussian squeezing in the presence of decoherence and imperfections \cite{Jeannic2018,Kala2022}.} The most common approach to introduce non-Gaussianity into light is via photon-resolving measurement in the Fock basis, performed on a part of an entangled state. This principle lies at the core of many state preparation protocols \cite{Konno_nlsq,KonnoGKP,Fadrn2024}. Certifying the successful preparation of non-Gaussian state is, however, complex, partly because of their fragility and partly because non-Gaussianity can manifest in many different forms \cite{Ogawa2016,Alexei,Asavanant17,Fadrn2024,KonnoGKP}. Properties shared by all non-Gaussian states, like Wigner negativity, often lack a straightforward operational interpretation in quantum information applications. The quality of an experimental realization of a specific target state can be evaluated via the excess noise of an operator whose ground state corresponds to the ideal target state \cite{Konno_nlsq,Kala2022,KonnoGKP,PetrPRL,Kala2025,Simon2025}. The non-Gaussianity is then certified by comparing the measured noise to the minimal value attainable by Gaussian states \cite{Miyata2016,Konno_nlsq,Kala2022,KonnoGKP}.

The successful implementation of continuous variable quantum computing then relies on embedding the non-Gaussian states into the cluster state. This is essential to avoid efficient classical simulation, which is otherwise possible for entirely Gaussian systems \cite{Chabaud2023}. To this end, photon subtraction or addition can be used to prepare multimode non-Gaussian states \cite{WalschaersPRLMMneg,Kala2025}. Furthermore, appropriate auxiliary non-Gaussian states, like the cubic phase state, can be entangled with or incorporated into the cluster state to unlock universality via measurement-induced nonlinear operations \cite{Miyata2016,Gu,Kalatelep}. Eventually, using GKP encoding of qubits into continuous-variable modes \cite{Gottesman2001} provides error correction against errors caused by finite squeezing in the cluster state \cite{LarsenFT}. All these approaches rely on entangling the cluster state with non-Gaussian states, i.e., preparing a non-Gaussian cluster state.

Here, we propose and develop nullifiers for non-Gaussian cluster states. These non-Gaussian nullifiers enable certification of non-Gaussianity directly within the cluster state and can be experimentally evaluated using standard homodyne measurements. To illustrate the concept, we also conduct a proof-of-concept experiment using photon-subtracted squeezed states, demonstrating the practical applicability of the proposed framework.

\section{Non-Gaussian nullifiers}
The canonical form of the Gaussian cluster state is prepared from eigenstates of the $p$ quadrature via continuous-variable $Cz$ gates \cite{CVcluster}. This two-mode gate transforms the $p$ quadratures as
    $p_1\rightarrow p_1 + x_2$ and
    $p_2\rightarrow p_2 + x_1$,
leaving the $x$ quadratures intact. The ideal Gaussian cluster state is found to be the ground state of the nullifiers
\begin{equation}\label{nullG}
    N_i = p_i - \sum_{j\in N(i)}x_j,
\end{equation}
which are a linear combination of quadrature operators. The summation goes over neighboring connected modes. The form is an inverse operation to the $Cz$ gates in the cluster state preparation and addresses the infinite squeezing of the $p$ quadrature eigenstates. An approximation created from finitely squeezed states then carries some noise in the nullifier operators.

In quantum optics, passive linear beam splitters are used instead of the $Cz$ gates as the latter require active online squeezing in their optical implementation \cite{Loock2007}. The preparation of the cluster state can be described in the Heisenberg picture as
\begin{equation}\label{symplectic}
       \textbf{r}'=V\textbf{r},
\end{equation}
where $\textbf{r}=(x_1,p_1,x_2,p_2...)^T$ is a set of quadratures corresponding to vacuum states, the symplectic transformation $V$ describes the action of the squeezers and $Cz$ gates, and $\textbf{r}'$ quadratures measured via homodyne measurement. The transformation $V$ can be decomposed via Bloch-Messiah decomposition into a sequence of a passive linear optical network, a set of squeezers, and another passive linear network \cite{Braunstein2005b,Loock2007}. As the input consists of vacuum states, the first linear optical network is irrelevant, and the whole cluster state preparation can be composed of a set of squeezers followed by a passive optical network with symplectic transformation $M$ \cite{Loock2007}. In this manner, the structure of the cluster state is created via a passive linear interferometer. Spatial-temporal multiplexing enables a compact implementation of a large cluster state generation \cite{MenicucciMultiplexing}.

\begin{figure}[ht]
\centering
\includegraphics[width=\linewidth]{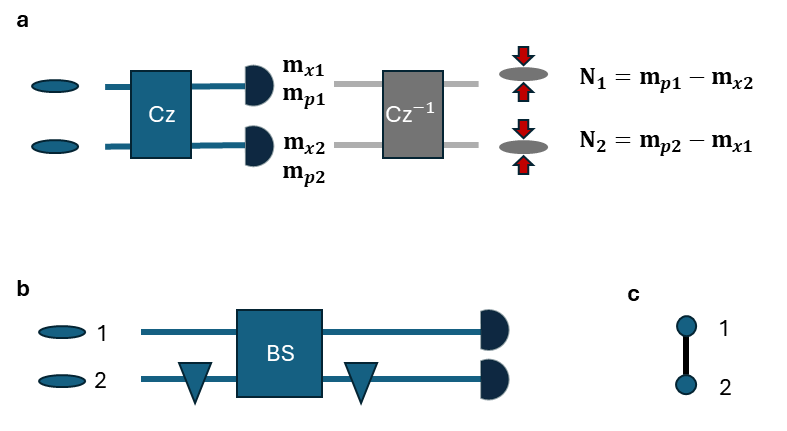}
\caption{\red{\textbf{a} Schematic illustration of a cluster state generation via $Cz$ gate and evaluation of the nullifiers $N_1$ and $N_2$ \eqref{2null} from homodyne measurement results $m_{x2},m_{p1}$ and $m_{x1},m_{p2}$, respectively. \textbf{b} Optical implementation using beam splitter and phase shifters. \textbf{c} Graph of the two-node cluster state.}
}
\label{nullifier}
\end{figure}

Let us describe a two-mode cluster state shown in Fig. \ref{nullifier} as an illustration. The cluster state has two nodes and a connecting edge. According to \eqref{nullG}, the nullifiers are of the form
\begin{equation}
\begin{split}
    N_1 =& p_1' - x_2'\\
    N_2=&p_2' -x_1'.
\end{split}
\end{equation}
The preparation of the cluster state starts with squeezed states with quadrature operators $x_1,p_1,x_2$ and $p_2$ that are connected to vacuum via $x_i=\exp(-r)x_i^{(0)}$ and $p_i=\exp(r)p_i^{(0)}$ for $i=1,2$. The passive linear optical network is shown in Fig. \ref{nullifier}\textbf{b}, consists of a phase shift applied to the second mode, a beam splitter interaction, and a second phase shift on the second mode. The overall transformation yields
\begin{equation}
    \begin{split}
        x_1'=& \frac{1}{\sqrt{2}}(x_1 + p_2),\;
        p_1'= \frac{1}{\sqrt{2}}(p_1 - x_2)\\
        x_2'=& \frac{1}{\sqrt{2}}(-x_2 - p_1),\;
        p_2' = \frac{1}{\sqrt{2}}(x_1-p_2).
    \end{split}
\end{equation}
This gives the nullifiers as
\begin{equation}\label{2null}
    \begin{split}
        N_1 = &\sqrt{2}p_1 = \sqrt{2}p_1^{(0)}\exp(r)\\
        N_2 = &\sqrt{2}p_2 = -\sqrt{2}p_2^{(0)}\exp(r),
    \end{split}
\end{equation}
thus vanishing as the initial squeezing goes to minus infinity, $r\rightarrow -\infty$

To reach universality and fault tolerance, it is necessary to embed non-Gaussian states into the Gaussian cluster state. In this case, the nullifiers of Gaussian cluster states can no longer be used for the cluster state certification in experimental data. We consider the following model of a cluster state. Some of the input modes are populated by a single-mode non-Gaussian state $\ket{\psi}_n$, the rest are initialized in the vacuum state. The cluster state is prepared from the set of initial states via Gaussian squeezers followed by a linear interferometer with unitary operation $U$ as
\begin{equation}\label{cluster}
    \ket{\Psi} = U (\otimes_n \ket{\psi}_n \otimes_m \ket{0}).
\end{equation}
Further, we define the characteristics of the individual states $\ket{\psi}_n$. Let $O_n$ be a positive semidefinite operator that has the state $\ket{\psi}_n$ as its nondegenerate ground state, i.e. a sole eigenstate of the lowest (zero) eigenvalue, and let it have a representation in terms of quadrature operators $x_n$ and $p_n$
\begin{equation}\label{null}
    O_n = O_n(x_n,p_n).
\end{equation}
Such a characterization was already developed for the cubic phase state \cite{Miyata2016,Brauer2021,Konno,Kala2022} and GKP states \cite{PetrPRL}. 

Using \eqref{null}, the $n_{th}$ non-Gaussian nullifier can be defined as
\begin{equation}\label{csnull}
    O_n = O_n(\{M^{-1}\textbf{r}'\}_{2n+1},\{M^{-1}\textbf{r}'\}_{2n+2}),
\end{equation}
where $\{\}_{2n+i}$ denotes the $n$-th row, addressing $x_{n+1}$ for $i=1$ and $p_{n+1}$ for $i=2$, and $M$ is a symplectic transformation equivalent to the evolution generated by the unitary operator U. The inverse symplectic transformation $M^{-1}$ enables us to untwist the cluster state structure \eqref{cluster} and access the statistics of the initial modes. Moreover, as the cluster was created by a linear operation, the "untwisting" can be performed directly on the measured data - a technique already used in the current verification of Gaussian cluster state preparation.

An experimentally prepared non-Gaussian cluster state $\rho_C$ \eqref{cluster} can deviate from the ideal one, either because of experimental imperfections or decoherence. Any alteration of the state results in a larger mean value of the nullifier $\textrm{Tr}[O_n\rho_C]$. This is due to the non-Gaussian nullifier having the ideal non-Gaussian cluster state as the nondegenerate ground state.  To be still able to say whether a non-Gaussian cluster state was successfully prepared beyond the Gaussian ones, we compare the nullifier mean value to the extremal value of the nullifier in Gaussian states and their mixtures. When the nullifier in $\rho_C$ exceeds this value, we can no longer say whether the state has any non-Gaussianity, at least within the framework of the non-Gaussian nullifiers. However, if the value is smaller than the Gaussian minimum, it verifies quantum non-Gaussianity in the cluster state, i.e. its incompatibility with any Gaussian state or mixture of Gaussian states. This is summarized in an inequality
\begin{equation}\label{neq}
    \textrm{Tr}[\rho_C O_n] < \min_G\textrm{Tr}[\rho_G O_n],
\end{equation}
where $\rho_G$ is a general Gaussian state, and the minimization is done over the set of Gaussian states and their mixtures.

The natural detection of optical cluster states is homodyne measurement. Its results obey statistics represented by generalized quadratures
\begin{equation}
    X(\theta) = \cos(\theta)x + \sin(\theta)p,
\end{equation}
where $\theta$ is the phase given by a local oscillator. Homodyne detection under two angles $\theta=0$ and $\theta=\frac{\pi}{2}$ enables evaluation of the noise in Gaussian nullifiers. For example the measurement results $m_{p_1}$ and $m_{x_2}$ obtained by measurement of $X_1(\frac{\pi}{2})$ and $X_2(0)$ can emulate measurement of the nullifier $p_1-x_2$ as $m_{p_1}-m_{x_2}$. The approach can be generalized for non-Gaussian nullifiers if the operators $O_n$ can be rewritten as a finite polynomial of the form 
\begin{equation}\label{4angles}
    \expval{O_n} = \sum_k c_k \expval{X(\theta_k)^{n_k}},
\end{equation}
which is a generalization of a technique previously used for the cubic nonlinear squeezing in optics \cite{Kala2022,Kala2025} and the cubic nonlinearity-based notion of nonclassicality for a mechanical oscillator \cite{Moore2019}.  The statistics of $X(\theta_k)$ is approximated via the data measured with the phase locked at $\theta_k$. From them, the expectation values in \eqref{4angles} are estimated. The form \eqref{4angles} does not require measurement of cross correlations such as $\expval{X(\theta_1)X(\theta_2)}$.

The expression \eqref{4angles} offers relative freedom in the choice of the operator $O_n$, as every Weyl symmetric term can be written in this manner
\begin{equation}\label{Wsym}
  :x^mp^n:_W  = \sum_{k=1}^{m+n} A_k X(\theta_k)^{m+n}.
\end{equation}
The proof can be found in appendix A.

\section{Example: Photon subtracted squeezed state} 
Let us apply the presented non-Gaussian nullifier to a specific example of a cluster state prepared from photon-subtracted squeezed states. The photon subtraction is a basic quantum operation, represented by an annihilation operator $a$ \cite{Wakui07, Neset2024}. This non-Hermitian operation is usually experimentally approximated in the following way. A small portion of the signal is reflected on a beam splitter with nearly unit transitivity and measured by a single photon detector. The realization is probabilistic, and the outcome is heralded by the successful separation of a single photon from the signal. The ideal photon-subtracted state can be rewritten as a squeezed single photon
\begin{equation}
    \ket{\psi}_n = \frac{1}{\mathcal{N}}aS(r)\ket{0}=S(r)\ket{1} ,
\end{equation}
where $S(r)$ is the Gaussian squeezing operator $S(r)=\exp(\frac{r}{2}(a^2-a^{\dagger 2}))$ and $\mathcal{N}=\sinh(r)^{-1}$.
Our aim is to define a non-Gaussian nullifier that characterizes a cluster state created from photon-subtracted Gaussian squeezed states. The first step is a choice of the operator $O_n$ \eqref{null}. The squeezed single photon state is found to be the ground state of the operator
\begin{equation}\label{kitsq}
    O_n = S(r)(n-1)^2S(-r).
\end{equation}
For $r=0$, $O_n$ consists of a number operator with shifted spectrum, such that the eigenvalue of the single photon Fock state $\ket{1}$ equals zero. The square then lifts all other eigenvalues to be positive. 

The operator $\eqref{kitsq}$ allows for a representation in quadrature operators \eqref{null} as well as a finite polynomial of the form \eqref{4angles}
\begin{equation}\label{angls}
\begin{split}
    O_n &= \frac{1}{4}[X(0)^4(g^4-\frac{1}{3}) + X(\frac{\pi}{2})^4(\frac{1}{g^4}-\frac{1}{3}) +  \\&   \frac{2}{3}(X(\frac{\pi}{4})^4 + X(\frac{-\pi}{4})^4) - 6(X(0)^2g^2 + X(\frac{\pi}{2})^2\frac{1}{g^2}) + 8],
\end{split}
\end{equation}
where $g$ denotes the transformation by squeezing $S(r)$ and $g = e^{-r}$. Thus, the nullifier is well suited for the characterization of cluster states measured by homodyne measurement. The choice of $\theta_k$ is not unique; similarly the angles $0, \frac{\pi}{6}, \frac{\pi}{2}$ and $\frac{5\pi}{6}$ can be used
\begin{equation}\label{expd}
    \begin{split}
        &O_n=S(r)\frac{1}{4}[\frac{8}{9}(X(\frac{\pi}{6})^4 + X(\frac{5\pi}{6})^4)  +\\& \frac{8}{9}X(\frac{\pi}{2})^4
           - 6 X(0)^2 - 6X(\frac{\pi}{2})^2 + 8]S^{\dagger}(r).
    \end{split}
\end{equation}

Next, we will find the extremal value of the nullifier in Gaussian states. Due to the linearity of trace, the minimum over Gaussian states will be attained for a pure state as
\begin{equation}\label{gmin}
    \min_G \textrm{Tr}[O_n\rho_G] \sim 0.611.
\end{equation}
The minimum was found via numerical optimization with a state
\begin{equation}
    \ket{\psi_G} = S(r')D(\alpha)\ket{0}.
\end{equation}
Once the expectation value of the nullifier \eqref{kitsq} in some state $\rho_C$ is below \eqref{gmin}, we have certified non-Gaussianity in the cluster state.

Another possible measurement to infer the nullifier mean value is from the heterodyne measurement. The measurement can be performed by splitting the signal on a beam splitter and measuring the two arms by $x$ and $p$ homodyne detections, respectively. Inevitably, vacuum fluctuations enter into play from the void port of the beam splitter. The resulting probability distribution is the Q-function and can be obtained by convolving the Wigner function with one unit of vacuum noise. The Q-function enables the evaluation of mean values as
\begin{equation}
    \expval{:O_n:_A} = \int \int d\alpha^2 O_n(\alpha,\bar{\alpha}) Q(\alpha,\bar{\alpha}),
\end{equation}
where the function $O_n(\alpha,\alpha^*)$ of complex variable $\alpha$ and its complex conjugate was obtained via replacing annihilation and creation operators $a\rightarrow\alpha,a^{\dagger}\rightarrow\bar{\alpha}$ in the antinormally ordered expression of $O_n$. In the case of the non-Gaussian nullifier \eqref{kitsq} presented here, the squeezing operation can be absorbed in the optimization of the Gaussian threshold state and the function of the complex amplitude $\alpha$ is found as
\begin{equation}
    O_n(\alpha,\bar{\alpha}) = |\alpha|^4 - 5|\alpha|^2 + 4.
\end{equation}

Due to the additional vacuum noise unit, the minimum over Gaussian states will increase when evaluated from heterodyne detection to
\begin{equation}
    \min_G \expval{:O_n:_A} \sim 0.75.
\end{equation}

Let us now analyze the behavior of the nullifier value in a more realistic scenario by considering an imperfect cluster state. The model is as follows: a photon subtraction is individually performed on a set of Gaussian squeezed states, then a Gaussian passive unitary $U$ is performed. We will consider two sources of imperfection. First, experimentally prepared squeezed states are usually not pure, meaning that the amount of antisqueezing is larger compared to the squeezing. Consequently, the states are not pure. Second, the successful splitting of a single photon from the signal during photon subtraction is heralded by measurement. In practice, the heralding detector possesses only limited quantum efficiency, which can be modeled by a sequence of a lossy channel with transitivity $\eta$ and an ideal detector \cite{Marek18}. Details can be found in Appendix B.

\begin{figure}[ht]
\centering
\includegraphics[width=1\linewidth]{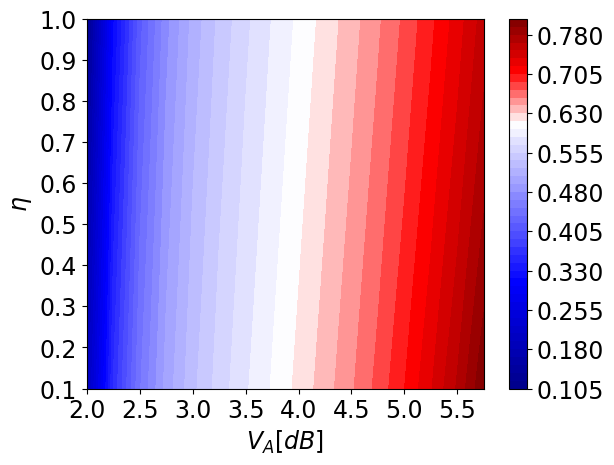}
\caption{The value of the non-Gaussian nullifier \eqref{kitsq} as a function of the quantum efficiency $\eta$ of the detector heralding the photon subtraction and as a function of antisqueezing $V_A$[dB] of the initial state, with squeezing fixed to -2 dB. The Gaussian threshold is indicated by the white stripe.
}
\label{etaN}
\end{figure}

The value of the non-Gaussian nullifier \eqref{kitsq} in the imperfect cluster state with varied initial antisqueezing and detector efficiency is shown in Fig. \ref{etaN}. A decrease in the efficiency of the detector imposes higher requirements on the purity of the initial squeezed state. The value of the antisqueezing has the leading effect, eventually causing the nullifier to increase over the Gaussian minimum as indicated by the white stripe. When the initial squeezing of -2 dB is accompanied by antisqueezing of 3.75 dB or higher (depending on the efficiency of the detector), the non-Gaussianity of the cluster state is no longer verifiable within the framework of the non-Gaussian nullifier.

\begin{figure}[ht]
\centering
\includegraphics[width=\linewidth]{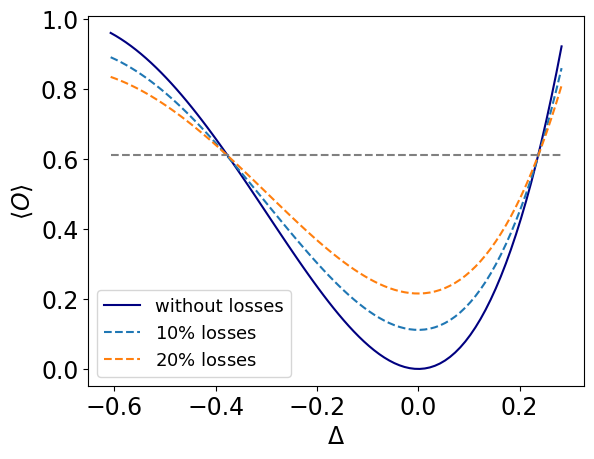}
\caption{Value of the non-Gaussian nullifier for photon-subtracted squeezed state cluster state, which was evaluated with imperfect knowledge of the cluster preparation. The state is prepared by a balanced beam splitter, whereas the nullifier is computed for a beam splitter with transitivity $t' = \frac{1}{\sqrt{2}} + \Delta$. The minimum can be seen when $\Delta=0$ and thus the linear operation in evaluation corresponds to an inverse of a balanced beam splitter. \red{Three cases are shown, an ideal cluster state and then a cluster state created from photon-subtracted squeezed vacua that were subject to $10\%$ and $20\%$ losses.}}
\label{uncU}
\end{figure}

Apart from the imperfections of the experimental setup, the quantum properties are also deteriorated by its inaccurate knowledge. As is schematically illustrated in Fig. \ref{nullifier}, the knowledge of the unitary transformation producing the considered cluster state is utilized in the data processing. Let us assume that the cluster state was prepared with a unitary operation $U$ from a set of non-Gaussian states. Additionally, let us assume that the equivalent of a different unitary transformation $U'$ was effectively applied to evaluate the non-Gaussian nullifier. Consequently, $U'^{\dagger}U$ does not yield an identity operator, and $U'\textrm{r}'U'^{\dagger}$ will not address the initial statistics of the non-Gaussian states. However, the operation $U'^{\dagger}U$, although not equal to identity, will be again a linear transformation equivalent to an interferometer. As such, it cannot imprint any non-Gaussianity and does not pose a risk of a false positive non-Gaussianity certification. As an illustration we include an example of the nullifier evaluation for a two-mode cluster state prepared from photon-subtracted squeezed vacua that interact on a balanced beam splitter, while in the data processing a beam splitter with different transitivity $t' = \frac{1}{\sqrt{2}}+ \Delta$ was considered. The results can be seen in Fig. \ref{uncU} and show the minimum of the non-Gaussian nullifier when evaluated in the true value of the transitivity and its deterioration for deviation from this value by $\Delta$.

\begin{figure}[ht]
\centering
\includegraphics[width=\linewidth]{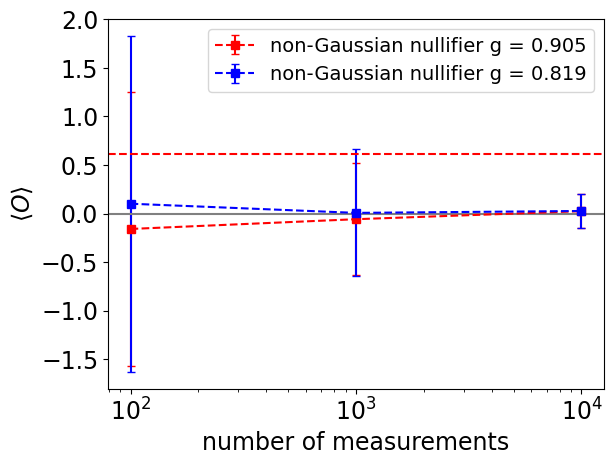}
\caption{Mean value and standard deviation of the non-Gaussian nullifier estimated from a given number of measurements per homodyne phase \red{simulated from statistics of an ideal photon-subtracted state}. The red line at $0.611$ depicts the Gaussian minimum of the nullifier. Red points correspond to initial squeezing given by $r=0.1$, blue points correspond to $r=0.2$.}
\label{Nmeas}
\end{figure}

The value of the non-Gaussian nullifier of an experimentally prepared cluster state can be estimated from homodyne measurement. When phase locking is available or if data corresponding to a fixed phase can be post-selected, the statistics of four generalized quadratures is sufficient in the discussed example of \eqref{angls}. The photon subtraction in the generation of the cluster state is probabilistic and thus requires stability of the experiment in time. Longer measurement time leads to datasets affected by experimental imperfections. To give an example, as the homodyne detection is phase sensitive, any phase drift affects its precision. Therefore, we analyze how the size of the dataset affects the precision of the nullifier estimation. We simulate random measurement values, which are drawn from an ideal theoretical probability of measuring $X(\theta)$ in an ideal cluster state. Data are further binned into $1000$ equi-width bins. In each bin, the mean value of simulated data points inside is computed together with their count, creating a histogram. The relative frequency of the mean value appearing given by the ratio of counts in a given bin and the overall number of data points then approximates the probability. Results from numerical simulation are shown in Fig. \ref{Nmeas} for $100 - 10 000$ measurements. The simulation was performed a hundred times to determine the spread of results caused by limited sampling from the probabilities. Ten thousand measurement points for each angle are sufficient to certify the non-Gaussianity in the cluster state.

\section{Experiment}
In order to illustrate the potential of the nullifier \eqref{kitsq} to capture non-Gaussianity in realistic states, we experimentally prepared the single photon state and squeezed single photon (``kitten'') states.

Kitten states at a wavelength of 1550 nm are generated by conditional photon subtraction from a squeezed vacuum state. The squeezed states are produced in an optical parametric oscillator (OPO) containing a 15 mm type-0 periodically poled KTP crystal in a bow-tie ring cavity, pumped at 775 nm by a frequency-doubled laser. The OPO has a bandwidth of 8 MHz, a threshold pump power of $\sim 800$ mW, and an escape efficiency of 97\%. For this experiment, the pump power was set to 50 mW.
 
Photon subtraction is implemented by tapping 3\% of the squeezed field on a highly asymmetric beam splitter and detecting a single photon in this arm. To ensure well-defined spectral mode selection, the tapped light passes through a narrow-band Fabry–Perot cavity (78 MHz bandwidth) before coupling into single-mode fiber, where it is further filtered by a 100 GHz DWDM filter, isolating the central degenerate mode at 1550.12 nm. Single-photon detection is performed using a fiber-coupled superconducting nanowire single-photon detector (SNSPD) with ~60\% detection efficiency and dark count rates of 50–100 Hz. The combined optical filtering and modulation stages yield an overall trigger channel efficiency of ~20\%. All optical cavities and measurement phases are actively stabilized using feedback control. During data acquisition, bright locking beams are momentarily blocked to prevent degradation of squeezing and saturation of the photon detector, employing a lock–measure cycle at a 100 Hz repetition rate.

The single photon state is prepared in a very similar way, with two main differences: The crystal is changed to one with type-2 phase matching, in order to generate polarisation non-degenerate photon pairs, and the output is split on a polarising beam-splitter instead of a 97/3 splitter. With this crystal, the pump threshold is rather high -- around 20 W -- so even with the pump power of 100 mW, the multi-pair production rate is negligible. 
 
The conditionally prepared states are characterized via time-resolved balanced homodyne detection. The homodyne detector combines a balanced beam splitter with a balanced detector (40 MHz bandwidth, 15 dB clearance) and achieves an overall detection efficiency of ~88\%. A strong local oscillator allows measurement of quadratures at controlled phases, with 10,000 traces acquired (5,000 for the single photon) at each 30$^\circ$ phase interval between 0$^\circ$ and 150$^\circ$. Each trace is recorded in a 1 µs window centered on the trigger click, sampled at 500 MS/s to resolve the OPO’s temporal correlations. Quadrature values are extracted by integrating each trace against a mode function matched to the OPO’s temporal response and the detection bandwidth. This choice maximizes the reconstructed Wigner function negativity. 
 
\begin{figure}[ht]
\centering
\includegraphics[width=0.8\linewidth]{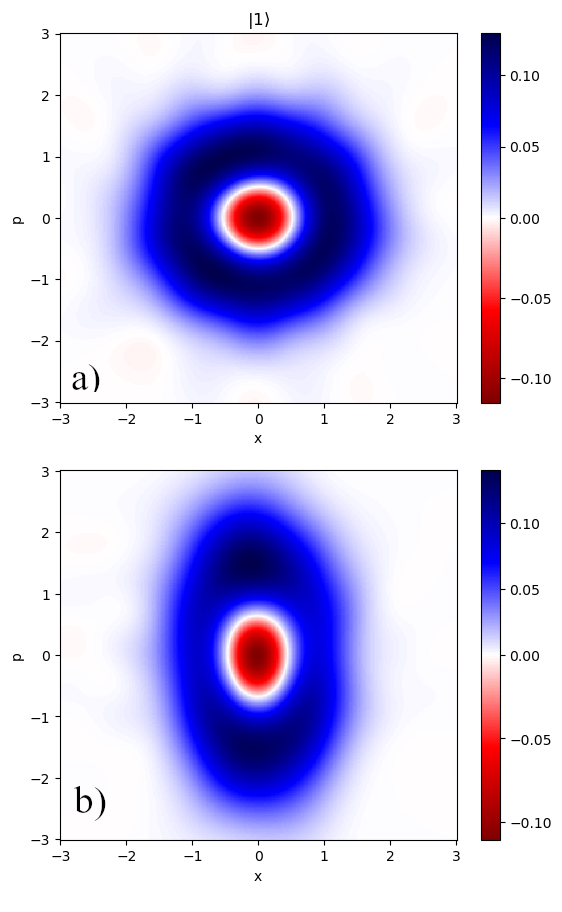}
\caption{Wigner functions of the experimentally prepared single photon state and squeezed single photon state.}
\label{Wigs}
\end{figure}

From the phase-resolved quadrature distributions, the density matrix and Wigner function of the generated state are reconstructed using a Maximum Likelihood algorithm. We do not correct for any losses or inefficiencies. This procedure yields the phase-space representation of the heralded non-Gaussian states, as illustrated in figure \ref{Wigs}. Both states have Wigner function minima of $W(0,0) = -0.11$.  

The expectation value of the nullifier was obtained with eq. \eqref{expd} as $0.3976(36)$ for $g=0.9752$ for the single photon state and $0.4810(20)$ for the squeezed single photon state with $g=1.6978$. Both states have the expectation value below the Gaussian threshold of $0.611$. The derivation of the uncertainty is described in more detail in Appendix C.

\section*{Conclusion}
We introduced the concept of non-Gaussian nullifiers as a practical tool for certifying the successful preparation of non-Gaussian cluster states. We showed that nullifiers constructed from Weyl symmetric operators can be directly evaluated from homodyne measurement, which is particularly advantageous as homodyne detection is the natural detection method for optical cluster states. As a concrete example, we considered a nullifier for a cluster state generated from photon-subtracted squeezed states and analyzed the impact of experimental imperfections, including imperfect photon subtraction, excess noise in the anti-squeezed quadrature, and finite precision in nullifier estimation. More generally, this framework extends Gaussian nullifier techniques into the non-Gaussian regime and provides a practical route toward verifying resource states for universal continuous-variable quantum computation. It is applicable to a wide class of non-Gaussian states, including cubic phase and GKP states, and offers a scalable alternative to full state tomography for future photonic quantum processors. Finally, we demonstrated experimentally how the proposed non-Gaussian nullifier can be used to identify photon-subtracted squeezed states.

\section*{Acknowledgement}
VK and PM acknowledge grant 25-17472S of Czech Science Foundation. JSNN, PM, RF, ULA and VK also acknolwedge Horizon Europe Research and Innovation Actions under Grant Agreement no. 101080173 (CLUSTEC). P.M. and R.F. acknowledge project CZ.02.01.010022\_0080004649 (QUEENTEC) of EU and the Czech Ministry of Education, Youth and Sport. JSNN, ULA, V.K. and R.F. acknowledges also the Quantera project ClusSTAR (8C24003) of the Czech Ministry of Education, Youth and Sport, the Innovation Fund Denmark and EU. JSNN and ULA acknowledge grant no.1063-00046A from Innovation Fund Denmark (PhotoQ). JSNN and ULA acknowledge project bigQ no. DNRF0142 from Danish National Research Foundation. JSNN and ULA acknowledge EU ERC project (ClusterQ no. 101055224). VK acknowledges project IGA-
PrF-2025-010 of the Palacký University. 
VK acknowledges support from the F.R.S.–FNRS under project CHEQS within the Excellence of Science (EOS) program.

\section*{Appendix A: Proof of eq. \eqref{Wsym}}
In order to prove the statement
\begin{equation}\label{AppA}
  :x^mp^n:_W  = \sum_{k=1}^{m+n} A_k X(\theta_k)^{m+n},
\end{equation}
we observe that an $n$-th power of generalized quadrature equals the following sum
\begin{equation}
    X(\theta)^n = \sum_{l=0}^n [\cos(\theta)]^{n-l}[\sin(\theta)]^l{n\choose l}:x^{n-l}p^l:_W.
\end{equation}
The sum has $n+1$ Weyl-symmetric terms. In order to separate one of them in the sense of \eqref{AppA}, we take $n$ different angles $\theta_k$. The coefficients $A_k$ can be found as a solution to a system of linear equations given by matrix
\begin{equation}
C = 
\begin{pmatrix}
\cos^n(\theta_1) & \cos^n(\theta_2) & \cdots \\
\cos^{n-1}(\theta_1)\sin(\theta_1) & \cos^{n-1}(\theta_2)\sin(\theta_2) & \cdots\\
\cos^{n-2}(\theta_1)\sin^{2}(\theta_1) & \cos^{n-2}(\theta_2)\sin^{2}(\theta_2) & \cdots\\
\vdots & \vdots & \ddots\\
\end{pmatrix},
\end{equation}
as
\begin{equation}
    C\textbf{A}=\textbf{d},
\end{equation}
where \textbf{A} is the vector of coefficients $A_k$ and \textbf{d} is a vector with $1$ on the position that determines which term we want to separate and otherwise equals zero. The system of linear equations has a solution, when the rank of $C$ equals the rank of the matrix with added right hand side $(C^T|\textbf{d})$. To this end it is sufficient to prove, that the matrix $C$ has a full rank. We start by multiplying the matrix $C$ by a diagonal matrix with diagonal
\begin{equation}
    E = \textrm{diag}(\frac{1}{\cos^n(\theta_1)},\frac{1}{\cos^n(\theta_2)},\frac{1}{\cos^n(\theta_3)}...)
\end{equation}
considering different $\theta_k$, this will change the value of the matrix $C$, however not its rank. The resulting matrix yields
\begin{equation}
CE = 
\begin{pmatrix}
1 & 1 & \cdots \\
\cot(\theta_1) & \cot(\theta_2) & \cdots\\
\cot^{2}(\theta_1) & \cot^{2}(\theta_2) & \cdots\\
\vdots & \vdots & \ddots\\
\end{pmatrix},
\end{equation}
and has the same rank as the matrix $C$. The rows are linearly independent and thus the matrix has a full rank.
\section*{Appendix B: Theoretical model of the imperfect photon subtracted squeezed state}
We introduce two sources of imperfections in the model of the photon subtracted squeezed vacuum. At first, we replace the initial squeezed vacuum with a mixed state. This is the usual situation in experiment, where for example losses can lead to a higher antisqueezing compared to the squeezing itself. Instead of defining the mixed state, it is possible to model stat with different squeezing and antisqueezing via losses applied to a pure squeezed vacuum. The covariance matrix of the squeezed vacuum yields
\begin{equation}
S_0 = 
\begin{pmatrix}
\frac{g}{2}&0 \\
0&\frac{1}{2g}\\
\end{pmatrix},
\end{equation}
after losses, that are modelled by a beam splitter with transitivity $t$ and reflectivity $r$, such that $t^2+r^2=1$, with the other mode being in vacuum state, the resulting density matrix is
\begin{equation}
S_0 = 
\begin{pmatrix}
t^2\frac{g}{2}+r^2\frac{1}{2}&0 \\
0&t^2\frac{1}{2g}+r^2\frac{1}{2}\\
\end{pmatrix}.
\end{equation}
Thus the losses enable parameterization of arbitrary squeezing and antisqueezing. The squeezing or antisqueezing in dB is defined as
\begin{equation}
    \nu = 10\log(\frac{V}{\frac{1}{2}}),
\end{equation}
where $V$ is the corresponding element of the covariance matrix. In our computation we model the state in Fock basis, defining the squeezing operation as
\begin{equation}
    S = \exp(\frac{r}{2}(a^2-a^{\dagger 2})),
\end{equation}
with
\begin{equation}
    r = \frac{\ln(g)}{2}
\end{equation}
and the beam splitter operation as
\begin{equation}
    U_{BS} = \exp(-\phi(a^{\dagger}b - ab^{\dagger})),
\end{equation}
where
\begin{equation}
    \phi = \arccos(t).
\end{equation}

Additionally, photon subtraction is usually provided by heralded splitting of a photon from the signal on a highly transmissive beam splitter. The heralding is done via an on/off detector, that only detects whether there is light present. This can be considered an approximation of the single photon measurement, when the intensity of the incoming light is low. The corresponding POVM element yields
\begin{equation}
    \Pi = 1 - \dyad{0}.
\end{equation}
Finite quantum efficiency of the detector $\eta$ can be simulated by additional losses before an ideal detection. Together with the losses, the corresponding POVM yields \cite{Marek18}
\begin{equation}
    \Pi' = 1 - \sum_{k=0}(1-\eta)^k\dyad{k}.
\end{equation}

\section{Appendix C: Precision estimation of the nullifier mean value derived from homodyne data}
The mean value of the nullified can be computed as
\begin{equation}
    \expval{O} = \sum_kc_k\expval{X(\theta_k)^{n_k}}.
\end{equation}
Having a homodyne measurement sample $\{X_i(\theta_k)\}$ of size $N$, the moments can be estimated as
\begin{equation}
    \expval{X(\theta_k)^{n_k}} = \frac{\sum_iX_i(\theta_k)^{n_k}}{N},
\end{equation}
the variance of the estimator can be found as
\begin{equation}
    \sigma^2_{k} = \frac{\sum_iX_i(\theta_k)^{2n_k}-\left(\expval{X(\theta_k)^{n_k}}\right)^2}{N-1}.
\end{equation}
In a similar manner covariances between potentially dependent moments is
\begin{equation}
\begin{split}
    &c_{kj} =\\ &\frac{(\sum_iX_i(\theta_k)^{n_k}-(\expval{X(\theta_k)^{n_k}}))(\sum_iX_i(\theta_j)^{n_j}-(\expval{X(\theta_j)^{n_j}}))}{N-1}. 
\end{split}
\end{equation}
The overall error can then be computed by
\begin{equation}
    \sigma^2 = \sum_k c_k^2\sigma^2_{k} + \sum_{kj}c_{kj}
\end{equation}

\bibliographystyle{unsrtnat}
\bibliography{references.bib}

\end{document}